\RequirePackage{ifpdf}
\ifpdf 
\documentclass[pdftex]{sigma}
\else
\documentclass{sigma}
\fi

\def\vq{{\bf q}}
\def\vr{{\br}}
\def\lf{\left(}
\def\rg{\right)}
\def\lq{\left[}
\def\rq{\right]}
\def\lgr{\left\{}
\def\rgr{\right\}}
\newcommand{\half}{{\scriptstyle{\frac{1}{2}}}}
\def\cA{{\cal A}}
\def\cE{{\cal E }}
\def\cR{{\cal R}}
\def\cT{{\cal T}}
\def\vA{{\bf A}}
\def\vB{{\bB}}

\def\bk{{\vec{k}}}
\def\bp{{\vec{p}}}
\def\vTheta{{\vec{\Theta}}}
\def\bTheta{{\mathbf \Theta}}
\def\bXi{{\mathbf \Xi}}
\def\bX{{\mathbf X}}
\def\by{{\vec{y}}}
\def\p{{\partial}}
\def\vx{{\vec x}}
\def\vX{{\vec X}}
\def\vb{{\vec b}}
\def\vP{{\vec P}}
\def\vp{{\vec p}}
\def\vQ{{\vec Q}}
\def\vk{{\vec k}}
\def\vnabla{{\vec\nabla}}

\def\nn{\nonumber}

\def\IR{{\mathbb{R}}} 

\def\smallover#1/#2{\hbox{$\textstyle\frac{#1}{#2}$}} %

\def\vq{{\vec q}}
\def\vp{{\vec p}}
\def\vP{{\vec P}}
\def\vQ{{\vec Q}}

\def\vq{{\vec q}}
\def\vr{{\vec r}}

\def\vx{{\vec x}}
\def\vA{{\vec A}}
\def\vB{{\vec B}}
\def\vnabla{{\vec\nabla}}

\numberwithin{equation}{section}

\begin{document}

\allowdisplaybreaks

\renewcommand{\thefootnote}{$\star$}

\renewcommand{\PaperNumber}{060}

\FirstPageHeading

\ShortArticleName{Exotic Galilean Symmetry and Non-Commutative
Mechanics}

\ArticleName{Exotic Galilean Symmetry \\ and Non-Commutative
Mechanics\footnote{This paper is a
contribution to the Special Issue ``Noncommutative Spaces and Fields''. The
full collection is available at
\href{http://www.emis.de/journals/SIGMA/noncommutative.html}{http://www.emis.de/journals/SIGMA/noncommutative.html}}}

\Author{Peter~A.~HORV\'ATHY~$^\dag$, Luigi MARTINA~$^\ddag$ and Peter C.~STICHEL~$^\S$}

\AuthorNameForHeading{P.A. Horv\'athy, L. Martina and P.C.~Stichel}

 \Address{$^\dag$~Laboratoire de Math\'ematiques et de Physique  Th\'eorique, Universit\'e de Tours,\\
 \hphantom{$^\dag$}~Parc de Grandmont, F-37200 Tours, France}
 \EmailD{\href{mailto:horvathy@lmpt.univ-tours.fr}{horvathy@lmpt.univ-tours.fr}}

\Address{$^\ddag$~Dipartimento di Fisica - Universit\`a del
Salento and Sezione INFN di Lecce,\\
\hphantom{$^\ddag$}~via Arnesano, CP. 193,
I-73100 Lecce, Italy}
\EmailD{\href{mailto:Luigi.Martina@le.infn.it}{Luigi.Martina@le.infn.it}}

\Address{$^\S$~An der Krebskuhle 21 D-33 619 Bielefeld, Germany}
\EmailD{\href{mailto:peter@physik.uni-bielefeld.de}{peter@physik.uni-bielefeld.de}}

\ArticleDates{Received March 23, 2010, in f\/inal form July 19, 2010;  Published online July 26, 2010}

\Abstract{Some aspects of the ``exotic'' particle, associated
with the two-para\-meter central extension of the planar Galilei
group are reviewed. A fundamental property is that it has
non-commuting position coordinates. Other and generalized
non-commutative
 models are also discussed.  Minimal as well as
anomalous coupling to an external electromagnetic f\/ield is
presented. Supersymmetric extension is also considered. Exotic
Galilean symmetry is also found in Moyal f\/ield theory. Similar
equations arise for a semiclassical Bloch electron, used to
explain the anomalous/spin/optical Hall ef\/fects.}


\Classification{46L55; 37K65; 37L20; 83C65; 81T70}

\renewcommand{\thefootnote}{\arabic{footnote}}
\setcounter{footnote}{0}

\section{Introduction: ``exotic'' Galilean symmetry}

A curious property of massive non-relativistic quantum systems is that Galilean boosts only act
up-to phase, so that only the $1$-parameter central extension
of the Galilei group  acts unitarily~\cite{LaLi}.
True representations only arise for massless
particles.

Do further extension parameters exist?  It is well-known that
central extensions are associated with non-trivial Lie algebra cohomology~\cite{Simms69,SSD}, and
Bargmann  \cite{Barg} has shown that, in $d\geq3$ space dimensions,
the Galilei group only admits a $1$-parameter central
extension, identif\/ied with the physical mass, $m$.
L\'evy-Leblond~\cite{LL} has
recognized, however, that,
owing to the Abelian nature of planar rotations,
 the \emph{planar} Galilei group admits a second central extension.
 The cohomology is two-dimensional, and is parametrized by
two constants, namely by the mass and a second,
 ``exotic'',  parameter $\kappa$. The second extension is highlighted by
the non-commutativity of Galilean boost generators,
\begin{gather*}
[K_1,K_2]=i\kappa.
\end{gather*}
 This fact has long been considered, however, a mere mathematical
curiosity, as
planar physics  itself has been viewed as a toy.

Around 1995 the situation started to change, though, with
the construction of physical models carrying such an ``exotic'' structure
 \cite{Grigore, otherNC,LSZ}.
\cite{LSZ} uses an acceleration-dependent
Lagrangian, while that in~\cite{DH}, is obtained following
Souriau's  method \cite{SSD}.

These models have the distinctive feature that the Poisson bracket
of the planar coordinates does not vanish,
\begin{gather}
\{x_1,x_2\}=-\frac{\kappa}{m^2}\equiv\theta,
\label{NCfreerel}
\end{gather}
and provide us with two-dimensional examples
of \emph{non-commutative mechanics} \cite{NaPo,NCMech,NCMech2,RoVe,Szabo,Delduc}\footnote{Our conventions are as follows. Symbols with arrows denote vectors, and those in boldface are tensors. The position vector~$\vr$ has coordinates~$x_i$.}.

What is the physical origin of exotic mechanics?
\emph{What is the quantum mechanical counterpart?} An
answer has been found soon after its introduction: it is a sort
of ``non-relativistic shadow'' of (fractional) spin~\cite{JaNa,anyons, anyoneqs}. Our particles can be interpreted
therefore, as nonrelativistic anyons \cite{HP1,HP2}.

The supersymmetric extension of the theory is outlined in Section~\ref{SUSY}.

All these examples have been taken from one-particle mechanics;
 exotic Galilean symmetry can be found, however,
 also in f\/ield theory~\cite{ExoFT}, as explained in Section~\ref{Moyal}.

Remarkably, similar
structures were considered,  independently and a\-round the same time, in condensed matter physics,
namely for the  Bloch electron \cite{Niu}:
 it was argued that the semiclassical dynamics
should involve a ``Berry term'', which induces
``anomalous'' velocity similar to the one in the
``exotic'' model \cite{DH}.

These, 3-space dimensional, models are also non-commutative, but the parameter $\theta$ is now promoted
to a  vector-valued  function of the
 quasi-momentum:  ${\vTheta}={\vTheta}(\bk)$.
Exotic Galilean symmetry, strictly linked to two space dimensions,
is lost.  However, a rich Poisson structure and an
intricate interplay with external magnetic f\/ields can be studied.
Further developments include the Anomalous \cite{AHE}, the Spin
\cite{SpinHall} and the Optical \cite{Optical,OptiHall,SpinOptics,OpticReview}
Hall ef\/fects.

This review completes and extends those in
\cite{Hreview,Mreview,Sreview}.

\goodbreak

\section{Exotic model, constructed by Souriau's method}\label{Sec2}

Models associated with a given symmetry group can be conveniently constructed using Souriau's method: the classical phase spaces of ``elementary systems'' correspond
to coadjoint orbits of their symmetry groups \cite{SSD}.
This was precisely
the way followed in~\cite{Grigore,DH} to produce an ``ele\-mentary'' classical system carrying L\'evy-Leblond's
``exotic'' Galilean symmetry. Translated from Souriau's to more standard terms, the model has an ``exotic''
symplectic form and a free Hamiltonian,
\begin{gather*}
\Omega_0 = dp_i\wedge dx_i+
\displaystyle\frac{1}{2}\theta \epsilon_{ij} dp_i\wedge{}dp_j,
\\
H_0 = \frac{\vec{p}{\,}^2}{2m}.
\end{gather*}
The associated free motions follow the usual straight lines,  described by the equations
\[
m \dot{x}_{i} =  p_{i}-m\theta\epsilon_{ij}\dot{p}_j, \qquad
 \dot{p}_{i} = 0.
 \]
  The ``exotic'' structure only enters the conserved
quantities, namely the boost and the angular momentum,
\begin{gather}
      j=\epsilon_{ij}x_ip_j+ \frac{\theta}{2} {\vec{p}}\,{}^2,
\qquad
K_{i}=-mx_{i}+p_{i}t-m\theta \epsilon_{ij}p_{j}.
\label{consquant}
\end{gather}
The ``exotic'' structure behaves hence
roughly as spin:
it contributes to some conserved quantities, but the
new terms are separately conserved.
The new structure does not
seem to lead to any new physics.

The situation changes dramatically, though, if the particle
is coupled to a gauge f\/ield.  Applying Souriau's coupling prescription
\cite{SSD} yields indeed
\begin{gather}
\Omega=\Omega_0+eB\,dq_1\wedge dq_2, \qquad H=H_0+eV,
\label{Sourcoup}
\end{gather}
where $B$ is the magnetic f\/ield assumed to be
perpendicular to the plane, and $V$ is the electric potential. For
simplicity, both $B$ and $V$ are assumed to be time-independent.

The associated Poisson bracket then automatically satisf\/ies the
Jacobi identity. The resulting equations of motion read
\begin{gather}
m^*\dot{x}_{i}
=
p_{i}-  em\theta\,\epsilon_{ij}E_{j},\qquad
\dot{p}_{i}
=
eE_{i}+eB\,\epsilon_{ij}\dot{x}_{j},
\label{DHeqmot}
\end{gather}
where the parameter $\theta=k/m^2$ measures the non-commutativity of $x_1$ and $x_2$, and
we have introduced the \textit{effective mass}
\begin{gather}
m^*=m(1-e\theta B).
\label{effmass}
\end{gather}

The novel features, crucial for physical applications, are
twofold: Firstly, the relation between velocity and momentum
contains an  ``anomalous velocity term''
$-  em\theta \epsilon_{ij}E_{j}$, so that
$\dot{x}_i$ and $p_{i}$ are not in general parallel. The second
one is the interplay between the exotic structure and the magnetic
f\/ield, yielding the ef\/fective mass $m^*$ in~(\ref{effmass}).

Equations (\ref{DHeqmot}) do not derive from a conf\/iguration-space
Lagrangian (but see Section \ref{Sec3}).  The $1$-order ``phase'' (in fact ``evolution space''~\cite{SSD}) formalism works,
however, so that the equations of motion (\ref{DHeqmot}) come from
varying the  action def\/ined by integrating the
``Cartan'' 1-form \cite{SSD},
\begin{gather*}
  \lambda = ({p_i}-{ A_i} ) d{ x_i}
-\frac{\vec{p}{\,}^2}{2m} dt + \frac{\theta}{2} \epsilon_{ij} {p_i} d{ p_j} 
\end{gather*}
along the lifted world-line $\widetilde{\gamma}$ in  evolution space $T^*{\IR}^2 \times {\IR}$. The
exterior derivative, $\sigma = d\lambda$, of the Cartan form
$\lambda$ provides us with a closed ``Lagrange--Souriau'' 2-form,
which, however, cannot be separated canonically into a
``symplectic'' and a ``Hamiltonian part''~\cite{SSD}. Thus  more
general procedures have to be adopted to build such a system and
clarify their Hamiltonian structure. These aspects will be
discussed in detail in the following  Sections~\ref{Sec3}, \ref{Sec4}
and \ref{Sec5}. Here we limit ourselves to notice that, in fact,
when $m^*\neq0$, (\ref{DHeqmot}) is a Hamiltonian system,
\begin{gather*}
\dot{\xi}=\{H,\xi^\alpha\},
\qquad \xi=(x_j,   p_i),
\end{gather*}
with Poisson brackets
\begin{gather}
\{x_{1},x_{2}\}=
\frac{m}{m^*} \theta,
\qquad
    \{x_{i},p_{j}\}= \frac{m}{m^*} \delta_{ij},
\qquad
    \{p_{1},p_{2}\}= \frac{m}{m^*} eB.
\label{exocommrel}
\end{gather}

A remarkable property is that for \textit{vanishing effective mass}
$m^*=0$, i.e., when the magnetic f\/ield takes the
critical value
\begin{gather*}
B=\frac{1}{e\theta},
\end{gather*}
the system becomes singular. Then ``Faddeev--Jackiw'' (alias
symplectic) reduction yields an  essentially two-dimensional,
simple system, reminiscent of ``Chern--Simons mechanics''~\cite{DJT}. The symplectic plane plays, simultaneously, the role
of both conf\/iguration and phase space. The only motions are those
which follow a generalized Hall law.

Quantization of the
reduced system yields, moreover, the ``Laughlin'' wave functions~\cite{QHE},
which are the ground states in the Fractional Quantum Hall Ef\/fect (FQHE).

The relations (\ref{exocommrel}) diverge as $m^*\to0$,
but after reduction we get, cf.~(\ref{NCfreerel}),
\begin{gather*}
\{x_1,x_2\}=\frac{1}{eB}=\theta.
\end{gather*}
The coordinates are, hence, non-commuting, and their
commutator is determined by the magnetic f\/ield, f\/ine-tuned to the parameter $\theta$.

{\bf Relation to another non-commutative mechanics.}
The exotic relations (\ref{exocommrel}) are similar to
those proposed (later) in~\cite{NaPo},
\begin{gather}
\{x_{i},x_{j}\}
   =\theta\epsilon_{ij},
\qquad
\{x_{i},p_{j}\}=\delta_{ij},
\qquad
\{p_{1},p_{2}\}=eB,
\label{NaPocommrel}
\end{gather}
which look indeed simpler. Using the standard Hamiltonian
$H=\frac{p^2}{2m}+eV(x)$, the associated equations of motion  read
\begin{gather}
mx_i' =p_i-em\theta\epsilon_{ij}E_j,
 \qquad
 p_i' =eB\epsilon_{ij} \frac{p_j}{m}+eE_i,
\label{NaPoeq}
\end{gather}
where we noted ``time'' by $T$ and  $(\cdot)'=\frac{d}{dT}$.

A short calculation reveals, however, that
\begin{gather}
\big\{x_i,\{p_1,p_2\}\big\}_{\rm cycl}=
e\theta\,\epsilon_{ij}{\partial}_jB, \nn
\end{gather}
so that \emph{the Jacobi identity is only satisfied if $B$ is a
constant}. In other words, the system (\ref{NaPocommrel}) is
\emph{only consistent} for a constant  magnetic f\/ield -- which is
an \emph{unphysical condition} in general\footnote{For another
role of the Jacobi identity in non-commutative mechanics, see
\cite{Chai}.}.

The model (\ref{NaPocommrel}) has another strange feature.
Let us indeed assume that the magnetic f\/ield is
radially symmetric, $B=B(r)$. One would then expect to have
conserved angular momentum. For constant $B$, applying
Noether's theorem to an inf\/initesimal  rotation
$
\delta \xi_i=\epsilon_{ij}\xi_j
$
yields indeed
$
\delta \xi_i=-\{j^{\rm NP},\xi_i\},
$
with
\begin{gather}
j^{\rm NP}= \frac{1}{1-e\theta B}
\underbrace{\left(\vx\times\vec{p}+
\frac{\theta}{2}\vec{p}\,{}^2+\frac{eB}{2}\vx\,{}^2
\right)}_{j} .
\label{JNangmom}
\end{gather}
This dif\/fers from the standard expression by the pre-factor
$(1-e\theta B)^{-1}.$ But what is even worse is that, for a radial but non constant magnetic f\/ield, (\ref{JNangmom}) is \emph{not conserved}:
\[
\frac{dj^{\rm NP}}{dT}=\frac{e\theta j}{(1-e\theta B)^2}
 \partial_iB x_i',
\]
while $j$ in (\ref{consquant})  is still conserved
as it should.

Can the theory def\/ined by (\ref{NaPocommrel}) be extended to an arbitrary $B$?

Let us f\/irst assume that $B=$ const.
 s.t.~$m^*\neq0$, and {\it let us redefine the time}\footnote{This was suggested to us by G.~Marmo (private communication).}, as
\begin{gather}
  T\to t=(1-e\theta B)T
  \quad\Rightarrow\quad
  \frac{\ d}{dT}=(1-e\theta B)  \frac{\ d}{dt} .
  \label{rescale}
\end{gather}

Then equations~(\ref{NaPoeq}) are carried into the exotic equations,
(\ref{DHeqmot}).
When $B={\rm const}$.
 s.t.~$m^*\neq0$, the two theories are therefore equivalent.

 Remarkably, the
\emph{time redefinition \eqref{rescale}} actually
\emph{extends the previous theory}, since it carries it
into the ``exotic model'', for which \emph{the Jacobi identity
 holds for any, not necessarily constant}~$B$. Thus, the transformation
(\ref{rescale}) (which is singular for $eB\theta=1$), removes the unphysical
restriction to constant magnetic f\/ields: (\ref{rescale})
regularizes the system~(\ref{NaPocommrel}).

\section{Acceleration-dependent Lagrangian in conf\/iguration space} \label{Sec3}

An independent and rather dif\/ferent
approach was followed in~\cite{LSZ}.

We start again with a particle characterized by the two central
charges
\[
m \qquad\hbox{and}\qquad
 \kappa=-m^2\theta
 \]
of the exotic Galilei group.
 These charges appear in the following Lie-brackets
 (represented by Poisson-brackets (PBs)) between the translation generators $P_i$ and the boost generators~$K_i$
\begin{gather}
\{ P_i, K_j \} = m\delta_{ij} , \qquad \{ K_i, K_j\} = - m^2\theta \epsilon_{ij}  .
\label{1}
\end{gather}
In order to f\/ind a conf\/iguration space Lagrangian whose Noether charges for boosts satisfy (\ref{1})
 we must add the second time derivative of the coordinates $\ddot{x}_i$ to the usual variables
 $x_i$ and $\dot{x}_i$. As shown in \cite{LSZ}
 the most general one-particle Lagrangian, which is at most linearly dependent on
  $\ddot{x}_i$, leading to the Euler--Lagrange equations of motion which are covariant
   w.r.t.\ the planar Galilei group, is given, up to gauge transformations, by
\begin{gather}
{\cal L} = \frac{m}{2} \dot{x}_i^2 + \frac{m^2\theta}{2} \epsilon_{ij} \dot{x}_i \ddot{x}_j  .
\label{2}
\end{gather}
Introducing the Lagrange multipliers $p_i$ and adding
$p_i(\dot{x}_i-y_i)$ to (\ref{2}) yields
\begin{gather}
{\cal L} = p_i \dot{x}_i +
\frac{m^2\theta}{2}\epsilon_{ij} y_i \dot{y}_j - H ({\by}, {\bp})
\label{3}
\end{gather}
with
\begin{gather*}
H({\by}, {\bp}) = y_i p_i - \frac{m}{2} y_i^2  .
\end{gather*}
(\ref{3}) describes a constrained system, because we have
\begin{gather*}
\frac{\partial {\cal L}}{\partial \dot{y}_i} = - \frac{m^2\theta}{2} \epsilon_{ij} y_j .
\end{gather*}
Therefore the PBs, obtained by means of the Faddeev--Jackiw procedure, take a non-standard form
\begin{gather}
\{ x_i, p_j\} = \delta_{ij}  , \qquad \{ y_i, y_j\} = - \frac{1}{m^2\theta} \epsilon_{ij}  .
\label{6}
\end{gather}
All other PBs vanish.

For the conserved boost generator we obtain
\begin{gather*}
K_i = - m x_i + p_i t - m^2\theta \epsilon_{ij} y_j
\end{gather*}
and therefore, due to (\ref{6}), the PB resp.~commutator of two boosts is nonvanishing
\begin{gather}
\{ K_i, K_j \} = - m^2\theta \epsilon_{ij}  .
\label{8}
\end{gather}
The Lagrangian~(\ref{3}) shows that the phase space is 6-dimensional. In order to split of\/f
 two internal degrees of freedom, we have to look for a Galilean invariant decomposition of the 6-dim phase space into two dynamically independent parts: a 4-dim external and a 2-dim internal part.
 This decomposition is achieved by the transformation \cite{LSZ,HP1} $ (\vx, \bp, \by) \to (\vX, \bp, \vQ)$
with
\begin{gather}
y_i = \frac{p_i}{m} + \frac{Q_i}{m \theta}
\qquad\hbox{and}\qquad
x_i = X_i - \epsilon_{ij}Q_j,
\label{9}
\end{gather}
leading to the following decomposition of the Lagrangian (\ref{3})
\[
{\cal L} = {\cal L}_{\rm ext} + {\cal L}_{\rm int}
\]
with
\begin{gather}
{\cal L}_{\rm ext} = p_i \dot{X}_i + \frac{\theta}{2} \epsilon_{ij} p_i \dot{p}_j - \frac{p_i^2}{2m}
\qquad\hbox{and}\qquad{\cal L}_{\rm int} = \frac{1}{2m\theta^2} Q^2_i + \frac{1}{2\theta} \epsilon_{ij} Q_i \dot{Q}_j  .
\label{10}
\end{gather}
From (\ref{9}) and the PBs (\ref{6}) it now follows that the new coordinates $X_i$ are noncommutative
\begin{gather}
\{ X_i, X_j \} = \theta\epsilon_{ij} .
\label{11}
\end{gather}
The remaining nonvanishing PBs are
\begin{gather*}
\{ X_i, p_j \} = \delta_{ij} , \qquad \{Q_i, Q_j \} = - \theta \epsilon_{ij}  .
\end{gather*}

\noindent
{\bf Conclusion.} The particle Lagrangian (\ref{2}) containing $\ddot{x}_i$ leads to
a nonvanishing commutator of two boosts. But in order to obtain noncommutative coordinates we are forced
 to decompose the 6-dim phase space in a Galilean invariant manner into
 two dynamically independent 4-dim external and 2-dim internal phase spaces.

{\bf Relation of the DH  and  LSZ models.}
The relation of the ``minimal''  and the con\-f\/i\-gu\-ra\-tion-space models of DH \cite{DH} and of L.S.Z.\ \cite{LSZ} respectively, has been studied in \cite{HP1}.
Introducing the coordinates $X_i$, $Q_i$,
and $p_i$ on $6$-dimensional phase space according to (\ref{9})
allows us to present
the symplectic structure and the
Hamiltonian associated with (\ref{2}) as
\begin{gather*}
 \Omega=dp_i\wedge dX_i
+
 \frac{\theta}{2}\varepsilon_{ij}dp_i\wedge dp_j
+
 \frac{1}{2\theta}\varepsilon_{ij}dQ_i\wedge
dQ_j,
\nonumber\\
H= \frac{\vp\,{}^2}{2m}- \frac{1}{2m
\theta^2}\vQ^2.
\end{gather*}
Thus,
 the model of L.S.Z.\ is decomposed into the
DH theory, augmented with a two-dimensional internal space,
and a negative zero point of the Hamiltonian.
Note that the  ``external'' and   ``internal'' phase spaces
 are ``almost'' independent: the only ef\/fect of internal motion
is indeed through the length of the internal vector, $|\vQ|$.

{\bf Generalization of (\ref{2}).} If we add to (\ref{2}) a term $f (\ddot{x}_i^2)$, the
obtained Lagrangian is the most general one involving, in a Galilean quasi-invariant manner, the variables $x_i$, $\dot{x}_i$ and $\ddot{x}_i$.

Then one can show\begin{description}\itemsep=0pt
\item{i)} the PB of the two boosts (\ref{8}) will not change,
\item{ii)} the new 8-dim phase space may be decomposed again in a Galilean invariant manner into two dynamically independent 4-dim parts, an external and an internal one.
\end{description}

{\bf Commutative -- versus noncommutative plane.}

The subalgebra of the Galilean algebra containing only
translations and boosts is given in the cases of, respectively,
their one- or two-fold central extensions by
\begin{alignat*}{3}
& \mbox{\underline{one-fold centrally extended}} \qquad && \mbox{\underline{two-fold centrally extended}}\\
&\{ p_i, K_j \} = m \delta_{ij}, \qquad &&\{ p_i^\prime, K_j^\prime \} = m \delta_{ij}, & \\
& \{ p_i, p_j\}  = 0,  && \{ p_i^\prime,  p^\prime_j \}   = 0, &\\
& \{ K_i, K_j \} = 0, && \{ K_i^\prime,  K_j^\prime \}  = - m^2\theta \epsilon_{ij}.&
\end{alignat*}

Obviously both are related by the transformations
\begin{gather*}
K^\prime_i = K_i - \frac{m\theta}{2} \epsilon_{ij} p_j, \qquad p_i^\prime = p_i .
\end{gather*}
To this corresponds the following point transformation between
noncommutative coordinates $X_i$ and commutative ones $q_i$
\begin{gather*}
X_i = q_i - \frac{\theta}{2} \epsilon_{ij} p_j
\end{gather*}
as can be read of\/f immediately from the form of ${\cal L}_{\rm ext}$ in (\ref{10}).

 Now the question arises: {\it What to use in physics,
the commutative or the non-commutative plane?}

{\bf Answer.} For free
particles both possibilities are equivalent.
 But in the case of a nontrivial interaction one has to use the commutative
(noncommutative) plane, if a local potential or gauge interaction
is given in terms of $q_i$ $(X_i)$.

It is worth mentioning that the acceleration-dependent
model presented in this Section can be related to
radiation damping~\cite{Neves}.

\section{General form of noncommutative mechanics} \label{Sec4}

Up to now noncommutativity has been described by a constant $\theta$ in the PB (\ref{11}). But it
 is possible to get $\theta$ as a function of ${\vX} $ and ${\bp}$  if one considers
external Lagrangians more general than~(\ref{10}).

To do this consider a very general class of Lagrangians given by
\begin{gather}
{\cal L} = p_i \dot{X}_i + \tilde{A}_i ({\vX}, {\bp}) \dot{p}_i - H({\bp}, {\vX})
\label{15}
\end{gather}
leading to the PBs
\begin{gather*}
\{ X_i, X_j \} \sim \epsilon_{ij} \tilde{B},
\qquad
\tilde{B} = \epsilon_{k\ell} \partial_{p_k} \tilde{A}_\ell ({\vX}, {\bp})
\end{gather*}
with
\[
\{ P_i, P_j\} = 0 .
\]
We dispense with the reproduction of the more complicated form of
the PBs for the phase space coordinates  $\lf X_i,
p_j\rg$.
 Again by the point transformation
\begin{gather*}
X_i \to q_i = X_i - \tilde{A}_i ({\vX}, {\bp})
\end{gather*}
we obtain commuting coordinates $q_i$ as follows from
\[
p_i \dot{X}_i + \tilde{A}_i \dot{p}_i = p_i \dot{q}_i + \frac{d}{dt} (\tilde{A}_i p_i)  .
\]

\bigskip
\noindent
\underline{Examples:}
\begin{description}\itemsep=0pt
\item{i)}
\begin{gather}
\tilde{A}_i = f(p^2) ({\vX} \cdot {\bp}\,) p_i
\label{18}
\end{gather}
leading to the PBs of the phase space variables
\begin{gather}
 \{X_i,X_j\} =
\frac{f(p^2)}{1-p^2 f(p^2)} \epsilon_{ij} L, \qquad L =
\epsilon_{k\ell} X_k p_\ell, \label{PB4}
\\
\{X_i,p_j\} = \delta_{ij}+\frac{f(p^2)}{1-p^2 f(p^2)}p_ip_j .
\label{19}
\end{gather}
A particular example is given by
\[
f(p^2) =  \frac{\theta}{1+p^2\theta}
\qquad\hbox{ and therefore} \qquad
 \frac{f}{1-p^2 f} =\theta.
\]
This gives exactly Snyder's NC-algebra, presented in 1947 \cite{Snyder}.

Another case, def\/ined by
\begin{gather}
f(p^2) = \frac{2}{p^2},
\label{20}
\end{gather}
can be related to a deformed Galilei algebra (to be discussed in
the next section).

\item{ii)}
\begin{gather*}
\tilde{A}_i = \tilde{A}_i ({\bp})
\end{gather*}
leading to the PBs
\begin{gather*}
\{ X_i, X_j\} = \epsilon_{ij} \tilde{B} ({\vec p}), \qquad \{ X_i,
p_j \} = \delta_{ij}  . 
\end{gather*}
$\tilde{B}$ is the Berry curvature for the semiclassical dynamics of electrons in condensed matter to be discussed in Section~\ref{Bloch}.

We may generalize \eqref{15} to the most general 1st-order Lagrangian
\begin{gather*}
{\cal L} = (p_i + A_i ({\vX}, {\bp})) \dot{X}_i + \tilde{A}_i ({\vX}, {\bp}) \dot{p}_i - H (\vp, {\vX}) .
\end{gather*}
Here $A_i ({\vX})$ describes standard electromagnetic interaction
(cp.\ Section~\ref{Sec6}, and Section~\ref{Bloch}
 for the 3 dimensional case, respectively). A particular case of a ${\bp}$-dependent $A_i$ has been
considered in~\cite{Ghosh06}.
\end{description}

\section[Lagrangian realization of the $\tilde{k}$-deformed Galilei algebra as a symmetry algebra]{Lagrangian realization of the $\boldsymbol{\tilde{k}}$-deformed Galilei algebra\\ as a symmetry algebra} \label{Sec5}

In 1991 Lukierski, Nowicki, Ruegg and Tolstoy invented
 the $k$-deformed Poincar\'{e} algebra \cite{LNRT}, which later found applications, e.g., in Quantum Gravity~\cite{GAC}. By rescaling the Poincar\'{e}
gene\-ra\-tors  and the deformation parameter $k$,  the
corresponding nonrelativisitic limit, the $\tilde{k}$-deformed
Galilei algebra, has been derived by Giller et al.~\cite{GKMMK}
and, in a dif\/ferent basis, by Azcarraga et al.~\cite{azcarraga}.
In this Section we will describe a Lagrangian realization of the
latter.

Again we look at the classical Lagrangian (\ref{15}) specif\/ied by (\ref{18}) and (\ref{20}) together with the
following choice of the Hamiltonian
\begin{gather}
H = \tilde{k} \ln (p^2/2) .
\label{24}
\end{gather}
According to (\ref{PB4}), (\ref{19}) we obtain the PBs
\begin{gather*}
\{ X_i, X_j \} = - \frac{2}{p^2} \epsilon_{ij} L \qquad \mbox{and} \qquad \{ X_i, p_j\} = \delta_{ij} - \frac{2}{p^2} p_i p_j 
\end{gather*}
which lead, together with the Hamiltonian (\ref{24}), to the
equations of motion
\begin{gather*}
\dot{p}_i = 0 \qquad \mbox{and}\qquad \dot{X}_i = - \frac{2\tilde{k}}{p^2}  p_i   .
\end{gather*}
Then,  we may def\/ine the ``pseudo-boosts'' $K_i$
\begin{gather*}
K_i = p_i t + \frac{p^2}{2\tilde{k}} X_i
\end{gather*}
which are conserved. They satisfy, together with $p_i$ and $H$,
the PB-algebra
\begin{gather}
\{ K_i, p_j\} = \frac{\delta_{ij}}{2\tilde{k}} p^2 - \frac{p_i
p_j}{\tilde{k}} ,
\qquad
\{ K_i, H\} = - p_i  ,
\qquad\{K_i, K_j\} = 0  .
\label{28}
\end{gather}
Together with the standard algebra of translations (represented by $p_i$ and $H$)
 and rotations (represented by $L$) the relations (\ref{28}), build the $\tilde{k}$-deformed Galilei algebra derived in \cite{azcarraga}.

The limit $\tilde{k} \to \infty$ leads to a divergent Hamiltonian (\ref{24}). Therefore, the $\tilde{k} $-deformation does not have a standard ``no-deformation limit''.

\section{Physical origin of the exotic structure} \label{Sec6}

A free  relativistic ``elementary'' particle in the plane
corresponds to a unitary representation of the planar Lorentz
group $O(2,1)$ \cite{anyons}. These representations are
in turn associated with the coadjoint orbits of $SO(2,1)$, endowed
with their canonical symplectic structures, interpreted by Souriau
as classical phase spaces~\cite{SSD}. Applied to the planar
Lorentz group, the procedure yields the \emph{relativistic} model
\cite{anyons, anyoneqs}
 \begin{gather*}
 \Omega_{\rm rel}= dp_\alpha\wedge dx^\alpha+
\frac{s}{2}\epsilon^{\alpha\beta\gamma}
 \frac{p_\alpha dp_\beta\wedge dp_\gamma}
 {(p^2)^{3/2}}  ,
 \\
 H_{\rm rel} = \frac{1}{2m}\big(p^2-m^2c^2\big).
 \end{gather*}
The $p$-dependent contribution looks like a
``magnetic monopole in momentum space'' (cf.~(\ref{pmonop}) below).

As pointed out by Jackiw and Nair \cite{JaNa},
 the free exotic model  can be recovered considering a
tricky non-relativistic  limit , namely
\begin{gather*}
 s/c^2\to \kappa=m^2\theta.
 \end{gather*}
The two-form $\Omega_{\rm rel}\big|_{H_{\rm rel}=0}$  goes indeed over
into the exotic symplectic form.
Intuitively, the exotic structure can be viewed as a
\emph{``non-relativistic
shadow'' of relativistic spin}.

The exotic Galilei group can itself be derived from the planar Poincar\'e group
 by  ``Jackiw--Nair''  contraction \cite{JaNa}. One starts with the  planar Lorentz generators,
\begin{gather*}
 \{J^\alpha,J^\beta\}=\epsilon^{\alpha\beta\gamma}J_{\gamma} .
 \end{gather*}
 For the classical system
 \begin{gather*}
 J_\mu=\epsilon_{\mu\nu\rho}x^\nu p^\rho+s\frac{p_\mu}{\sqrt{p^2}} .
 \end{gather*}
A non-relativistic boost is the ``JN'' limit of
a suitable Lorentz transformation,
 \begin{gather*}
 \frac{1}{c}\epsilon_{ij}J^j
 \to mx_i-p_it+m\theta\epsilon_{ij}p_j=-K_i,
 \end{gather*}
and the  exotic relation is recovered,
\begin{gather*}
\{K_1,K_2\}=J_0/c^2\to
 \frac{s}{c^2}=\kappa.
 \end{gather*}
The angular momentum is in turn
 \begin{gather*}
 J_0=\vec{x}\times\vec{p}+s+\frac{s}{m^2c^2}\vec{p}\,{}^2
 \to \vec{x}\times\vec{p}+\frac{\theta}{2}\vec{p}\,{}^2=j .
 \end{gather*}
whereas the divergent term $s=\kappa c^2$ has to be
removed by hand.

It is worth mentioning that the ``Jackiw--Nair limit'' of a relativistic particle with torsion \cite{MPtor}
provides us with the L.S.Z.\ model \cite{HP1}, and a similar procedure yields the so-called ``Newton--Hooke''
system \cite{PlNewtonHooke}.
Applied to the inf\/inite-component Majorana-type  anyon equations \cite{anyons,anyoneqs}
 yields, furthermore, a  f\/irst-order inf\/inite-component ``L\'evy-Leblond type''  system with exotic Galilean symmetry~\cite{HP2}.

About anyons constructed from orbits, see also~\cite{Negro}.

\section{Anomalous coupling of anyons} \label{Sec7}

It has been suggested \cite{CNP}  that
a classical, \emph{relativistic} anyon in an electromagnetic f\/ield should be described by the equations
\begin{alignat}{3}
& m \frac{dx^\alpha}{d\tau} = p^\alpha \qquad
&&\hbox{(velocity-momentum)}, &\nonumber\\
& \frac{dp^\alpha}{d\tau}=\frac{e}{m}F^{\alpha\beta}p_\beta
 \qquad &&\hbox{(Lorentz equation)}, &\label{CNP}
\end{alignat}
$\alpha,\beta,\ldots = 0,1,2$ and $\tau$ denotes the proper time.
These equations are Hamiltonian, with symplectic form and
Hamilton's function
\begin{gather}
 \Omega = \Omega_{\rm rel}+\frac{1}{2} eF_{\alpha\beta}dx^\alpha\wedge dx^\beta,
\nonumber \\
 H = H_{\rm rel}+  \frac{es}{2m\sqrt{p^2}}\epsilon_{\alpha\beta\gamma}
 F^{\alpha\beta}p^\gamma,
 \label{CNPHam}
 \end{gather}
respectively. Let us observe that the second, non-minimal term
 in the Hamiltonian is \emph{dictated by the required form of
the velocity relation} in~(\ref{CNP}).

The
model of~\cite{CNP} has gyromagnetic ratio $g=2$,
and some theoreticians  have long  believed~\cite{CNP,anyong} that this is indeed
the ``correct'' $g$ value of anyons. Experimental evidence
 shows, however,  that in the Fractional Quantum Hall Ef\/fect, for example,
the measured value of $g$ is approximately \emph{zero}~\cite{g0}.
Is it possible to construct an ``anomalous'' model with
$g\neq2$? The answer is af\/f\/irmative \cite{AnAn}, as we now explain.

  Planar spin has to satisfy the relation $S_{\alpha\beta}p^\beta=0$.
The spin tensor has, therefore, the form
\begin{gather*}
S_{\alpha\beta}=
\frac{s}{\sqrt{p^2}}\epsilon_{\alpha\beta\gamma}p^\gamma.
\end{gather*}
Introducing the shorthand  $-F_{\alpha\beta}S^{\alpha\beta}=F\cdot S$,
the Hamiltonian  (\ref{CNPHam}) is presented as
\begin{gather}
H^{\rm CNP}=\frac{1}{2m}\big(p^2-M^2c^2\big)
\qquad\hbox{where}\qquad
 M^2=m^2+\frac{e}{c^2}F\cdot S.
 \label{CNPHambis}
\end{gather}

Let us observe that the ``mass'' $M$ depends here on spin-f\/ield coupling.
Our clue for gene\-ra\-li\-zing this model has been the formula put forward
by Duval  \cite{DThese,Souriau74}: let us posit, instead
of~(\ref{CNPHambis}), the mass formula
\begin{gather*}
 M^2=m^2+ \frac{g}{2} \frac{e}{c^2}F\cdot S,
\end{gather*}
where $g$ is an arbitrary real constant. Then {\it consistent
equations of motion} are obtained
 for any~$g$, namely
\begin{gather}
D\frac{dx^\alpha}{d\tau} =
G\frac{p^\alpha}{M}+(g-2) \frac{es}{4M^2}
\epsilon^{\alpha\beta\gamma}F_{\beta\gamma},
\label{DHvel}
 \\
\frac{dp^\alpha}{d\tau}= \frac{e}{m}F^{\alpha\beta}p_\beta,
\label{DHLorentz}
\end{gather}
where the coef\/f\/icients denote the complicated, f\/ield-dependent expressions
\begin{gather*}
D=1+\frac{eF\cdot S}{2M^2c^2},
\qquad
 G=1+\frac{g}{2} \frac{eF\cdot S}{2M^2c^2}.
\end{gather*}
\emph{Choosing} $g=2$, the generalized model plainly reduces to
equation~(\ref{CNP}), proposed in~\cite{CNP}. We stress, however,
that \emph{no physical principle requires that the second,
``anomalous'' term should vanish} in (\ref{DHvel}). $g=2$ is
\emph{not} a physical necessity therefore: a perfectly consistent
model is obtained for any $g$, as it has been advocated a long
time ago~\cite{DThese,Souriau74}.

{\bf Non-relativistic anyon with anomalous coupling.}
We can now consider the ``Jackiw--Nair'' non-relativistic limit of the above
relativistic model. This provides us, for  any $g$, with the
Lorentz equation~(\ref{DHLorentz}), supplemented with
\begin{gather*}
(M_gD)\dot{x}_i=
Gp_i-\left(1-\frac{g}{2}\right)
eM_g\theta\epsilon_{ij}E_j,
\end{gather*}
where
\begin{gather*}M_g=m(\sqrt{1-g\theta eB}),
\qquad
D=\big(1-(g+1)\theta eB\big),
\qquad
G=\big(1-(3g/2))\theta eB\big).
\end{gather*}

$\bullet$ It is a most important fact that,
for {\it any} $g\neq2$, the only consistent motions follow a~gene\-ra\-li\-zed Hall law, whenever the f\/ield takes {\it either}
of the critical values
\begin{gather*}
B=\frac{1}{1+g} \frac{1}{e\theta}
\qquad\hbox{or}\qquad
\frac{2}{3g} \frac{1}{e\theta}.
\end{gather*}
One can indeed show that, for any $g\neq2$, the models can
be transformed into each other by a~suitable redef\/inition.
For $g=0$ the equations become identically satisf\/ied.
See~\cite{AnAn} for details.

$\bullet$ In particular, for $g=0$ the minimal exotic  model of~\cite{DH}
is recovered. The latter is, hence, {\it not} the  NR limit of the model of~\cite{CNP} (\ref{CNP}) [which has $g=2$, as said].
\emph{The experimental evidence~{\rm \cite{g0}} is, hence, a strong argument
in favor of the minimal model of {\rm \cite{DH}}}.

$\bullet$ $g=2$ is  the only case when
the velocity  and  the momentum are parallel. This is, however,
{\it not} required by any f\/irst principle.

Having  an anomalous velocity  relation
seems to be unusual in high-energy physics; it is, however, a well accepted
 requirement in condensed matter physics, as explained
in Section~\ref{Bloch}.

Let us mention that relativistic anyons can be described,
at the f\/ield theoretical level, by
inf\/inite-component f\/ields of the  Majorana--Dirac type~\cite{anyoneqs}.
Coupling them
 to an external gauge f\/ield is a major unsolved problem. Partial results can be obtained in the non-relativistic case~\cite{HP3}.

\section{Two ways of introducing electromagnetic interactions} \label{Sec8}

In this section we will show that Souriau's coupling prescription~(\ref{Sourcoup}) is not the only possibility to introduce electromagnetic (e.m.)
 interaction into the Lagrangian ${\cal L}_{\rm ext}$~(\ref{10}).

In the commutative case we have the principle of minimal e.m.
coupling
\begin{gather*}
p_i \dot{X}_i - \frac{p^2_i}{2m} \to (p_i + eA_i ({\vX}, t)) \dot{X}_i
- \frac{p_i^2}{2m} + eA_0 ({\vX}, t),
\end{gather*}
called the minimal additon rule, which is
equivalent, due to the point transformation $p_i \to p_i - e A_i$,
to the minimal substitution rule  \cite{stichel2},
\begin{gather*}
p_i \dot{X}_i - \frac{p_i^2}{2m} \to p_i \dot{X}_i - \frac{(p_i - eA_i)^2}{2m} + e A_0 ({\vX}, t)  .
\end{gather*}
In the noncommutative case the equivalence of minimal addition and minimal substitution rule is not valid.
Therefore we have to consider two dif\/ferent ways of introducing the minimal e.m. coupling:

{\bf Minimal addition} (Duval--Horvathy \cite{DH}, called \textit{DH-model})
\begin{gather}
{\cal L}  \to {\cal L}_{\rm e.m.} = {\cal L} + e(A_i \dot{X}_i + A_0),
\label{minadd}
\end{gather}
which, as usual, is quasi-invariant w.r.t.\ standard gauge
transformations
\begin{gather*}
A_\mu ({\vX}, t) \to A_\mu ({\vX},t) + \partial_\mu \Lambda ({\vX}, t)  .
\end{gather*}

Obviously the minimal addition rule (\ref{minadd}) is equivalent to Souriau's prescription (\ref{Sourcoup}).

{\bf Minimal substitution}
(Lukierski--Stichel--Zakrzewski \cite{stichel2}, called
{\it L.S.Z.\ model})\footnote{ In this model the gauge
f\/ields carry a ``hat'' in order to distinguish them from the
corresponding quantities in the DH-model.}
\begin{gather*}
H = \frac{p^2_i}{2m} \to
H_{\rm e.m.}=\frac{(p_i-e\hat{A}_i)^2}{2m}-e\hat{A}_0 .
\end{gather*}
The corresponding Lagrangian is quasi-invariant w.r.t.\ generalized
gauge transformations, given in inf\/initesimal form by
\begin{gather*}
\delta \hat{A}_\mu ({\vX},t) = \hat{A}_\mu^\prime ({\vX} + \delta
{\vX}, t)- \hat{A}_\mu ({\vX}, t) = \partial_\mu \Lambda ({\vX},
t),
\end{gather*}
with
\begin{gather}
\delta X_i = - e \theta \epsilon_{ij} \partial_j \Lambda
\label{8.7}
\end{gather}
and supplemented by
\begin{gather*}
\delta p_i = e \partial_i \Lambda .
\end{gather*}

Note that the coordinate transformations (\ref{8.7}) are area preserving.

It turns out that both models are related to each other
by a noncanonical transformation of phase space variables supplemented by
 a classical Seiberg--Witten transformation of the corresponding gauge potentials:

If we denote the phase space variables and potentials for
\begin{enumerate}\itemsep=0pt
\item[--] the DH-model by $(
\vec{\eta}, \vec{{\cal P}}, A_\mu)$,

\item[--] the L.S.Z.-model by $({\vX}, {\bp}, \hat{A}_\mu)$,
\end{enumerate}
\noindent then we f\/ind the relations
\begin{gather*}
\eta_i ({\vX},t) = X_i + e\theta \epsilon_{ij} \hat{A}_j ({\vX}, t),
\\
{\cal P}_i = p_i - e \hat{A}_i ({\vX},t)
\end{gather*}
with the corresponding f\/ield strengths related by
\begin{gather}
\hat{F}_{\mu \nu} ({\vX},t) = \frac{F_{\mu \nu} (\vec{\eta}, t)}{1-e\theta B ( \vec{\eta},t)}  .
\label{8.11}
\end{gather}
The Seiberg--Witten transformation between the resp.
gauge f\/ields is more involved,  and will not be reproduced here
(for details cp.~\cite{stichel2}).

These results lead to an interesting by-product:
Consider the PBs of coordinates in both models, given by
\begin{gather}
\{ \eta_i, \eta_j\} = \frac{\theta\ \epsilon_{ij}}{1-e\theta B(
\vec{\eta},t)} \qquad\mbox{and} \qquad \{ X_i, X_j\} =
\theta\, \epsilon_{ij}. \label{PB8}
\end{gather}
Then the foregoing results implicitly give the coordinate
transformation between a~model with a~constant noncommutativity parameter $m^2\theta$ and
one with  arbit\-ra\-ry coordinate-dependent noncommutativity function $m^2\theta ({\vX},t)$ (this result has been rediscovered  in~\cite{fosco}).

Now the question arises, which of both models has to be used for physical applications?
Let us look at one example, the Quantum Hall ef\/fect.
As already shown in Section~\ref{Sec2}
in the case of the DH-model~\cite{DH} the Hall law,
\begin{gather}
\dot{X}_i = \epsilon_{ij} \frac{E_j}{B},
\label{HallLaw}
\end{gather}
is valid at the critical magnetic f\/ield
\begin{gather*}
B_{\rm crit} = \big(e\theta\big)^{-1} .
\end{gather*}
Then it follows from the f\/ield transformation law (\ref{8.11})
that, for the L.S.Z.-model, the Hall law is valid in the limit of
large e.m.\ f\/ields. In order to
see this in more detail we have to consider the equations
of motion for the L.S.Z.-model formulated in terms of the
gauge-invariant phase space variables $\vec{\eta}$ and
$\vec{\cal P}$. For that, we use the
equations of motion (\ref{DHeqmot}) for the DH model
written in terms of $\eta_i$ and ${\cal P}_i$, transform the e.m.\
f\/ields according to (\ref{8.11}) and we obtain $(e=1$, $m=1)$
\begin{gather}
\dot{\eta}_i  =  (1+\theta\hat{B}){\cal P}_i-\theta \epsilon_{ij} \hat{E}_j,
\nonumber\\
\dot{{\cal P}}_i  =  \hat{B} \epsilon_{ij} {\cal P}_j + \hat{E}_i . \label{815}
\end{gather}
For the particular case of homogeneous e.m.\ f\/ields we obtain f\/inally
\begin{gather}
\ddot{\eta}_i = \hat{B} \epsilon_{ij} \dot{\eta}_j + \hat{E}_i
\label{Hallbis}
\end{gather}
leading to the Hall law (\ref{HallLaw}) in the high f\/ield limit.

Note that~(\ref{Hallbis}) has the same functional form as in the commutative case.

Another point of view is presented in~\cite{Acatrinei2003}.

\section{Supersymmetry}\label{SUSY}

In the following, we supersymmetrize the e.m.~coupling models
treated in the last section.
 To do that we follow the treatment in Section~3 of \cite{stichel3}. For that, we
consider standard $N=2$ SUSY characterized by
\begin{gather}
H = \frac{i}{2} \{ Q, \bar{Q}\}
\label{QuantHam}
\end{gather}
and
\begin{gather}
\{ Q, Q\} = \{ \bar{Q}, \bar{Q}\} = 0  .
\label{9.2}
\end{gather}
In order to construct the supercharge $Q$, satisfying~(\ref{QuantHam}), we start with the common structure of the bosonic Hamiltonian $H_b $ for both models $(e=1$, $m=1)$
\begin{gather}
H_b = \frac{1}{2} \big({\cal P}^2_i + W_i^2 ({\vX})\big)
\label{9.3}
\end{gather}
with
\[
{\cal P}_i = p_i \qquad \mbox{for the DH-model}
\]
and
\[
{\cal P}_i = p_i - A_i \qquad \mbox{for the L.S.Z.-model}.
\]
Note that, in accordance with the quantized form of~(\ref{QuantHam}), the potential term in (\ref{9.3}) is chosen to be positive
\begin{gather}
A_0 = - \frac{1}{2} W^2_i .
\label{9.4}
\end{gather}
In order to add to (\ref{9.3}) its fermionic superpartner, we supplement the bosonic
 phase space variables with fermionic coordinates $\psi_i (\bar{\psi}_i)$ satisfying canonical PBs
\begin{gather*}
\{ \psi_i, \bar{\psi}_j \} = - i \delta_{ij} .
\end{gather*}
Now we assume
\begin{gather*}
Q = i ({\cal P}_i + i W_i)\psi_i
\end{gather*}
such that (\ref{9.3}) is valid. But now the relations (\ref{9.2}) are fulf\/illed only if the following two
conditions are satisf\/ied:
\begin{gather}
\{ {\cal P}_i , {\cal P}_j\} = \{ W_i, W_j\}
\label{9.7}
\end{gather}
and
\begin{gather}
\{ {\cal P}_i, W_j\} = \{ {\cal P}_j, W_i \} .
\label{9.8}
\end{gather}
It can be shown that (\ref{9.8}) is satisf\/ied automatically in both models, whereas~(\ref{9.7}) f\/ixes the magnetic f\/ield in terms of~$W_i$ (same form for both models):
\begin{gather}
B = \frac{\theta}{2} \epsilon_{ij} \epsilon_{k\ell} \partial_k W_i \partial_\ell W_j  .
\label{9.9}
\end{gather}
The connection between B-f\/ield (\ref{9.9}) and electric potential $A_0$~(\ref{9.4}) takes a simple form in the case of rotational invariance. From
\begin{gather*}
W_i ({\vX}) = \partial_i W (r)
\end{gather*}
we obtain
\begin{gather*}
A_0 (r) = - \frac{1}{2} (W^\prime (r))^2
\end{gather*}
and
\begin{gather*}
B(r) = - \frac{\theta}{r} A^\prime_0 (r) .
\end{gather*}
As an example, consider the harmonic oscillator.
Then
\begin{gather*}
A_0 = - \frac{\omega^2}{2} r^2
\end{gather*}
and we obtain a homogeneous $B$-f\/ield of strength
\begin{gather*}
B = \theta \omega^2  .
\end{gather*}

The supersymmetric extension of the DH model, and of anyons,
 have also been studied in~\cite{ACHP} and in~\cite{anyonsusy}, respectively.

\section{Galilean symmetry in Moyal f\/ield theory}\label{Moyal}

As we mentioned already, the physical explanation of the Fractional Quantum Hall Ef\/fect (FQHE) relies on the dynamics of
quasiparticles which carry both an electric and a magnetic charge~\cite{QHE}. In the f\/ield theory context,
these quasiparticles arise as charged vortex solutions of the
coupled f\/ield equations. The phenomenologically
preferred theory of Zhang et al.~\cite{ZHK} is Galilei invariant; the
 Galilean boost commute for these models, though.
Does there exist a f\/ield theoretical model with
 ``exotic'' Galilean symmetry? The answer is yes, if we
consider Moyal f\/ield theory~\cite{Szabo,NCFT}. Here one
considers the usually-looking Lagrangian
\begin{gather*}
    L=
    i\bar{\psi} D_{t}\psi-\frac{1}{2}\big|{\vec{D}\psi}\big|^2
    +\kappa\left(\frac{1}{2}\epsilon_{ij}
    \partial_{t}A_{i}A_{j}+A_{t}B\right)
 \end{gather*}
but where the covariant derivative and the f\/ield
strength,
\begin{gather*}
    D_{\mu}\psi=\p_{\mu}\psi-ieA_{\mu}\star\psi,
    \\
    F_{\mu\nu}=\p_{\mu}A_{\nu}-\p_{\nu}A_{\mu}
    -ie\big(A_{\mu}\star A_{\nu}-A_{\nu}\star A_{\mu}\big),
\end{gather*}
respectively, involve  the Moyal ``star'' product, associated
with the parameter $\theta$,
\begin{gather*}
\big(f\star g\big)(x_1, x_2)=\exp\left(i\frac{\theta}{2}\big(
\p_{x_1}\p_{y_2}-\p_{x_2}\p_{y_1}\big)\right)
f(x_1, x_2)g(y_1, y_2)\Big|_{\vx=\vec{y}}.
\end{gather*}
Here the matter f\/ield $\psi$
is in the fundamental representation of the gauge group
$U(1)_{*}$ i.e., $A_{\mu}$~acts from the left.
The associated f\/ield equations look formally
as in the commutative case,
\begin{gather}
    iD_{t}\psi+\frac{1}{2}\vec{D}^2\psi = 0,\nonumber
    \\
   \kappa E_{i}-{e}\epsilon_{ik}j^{l}_{\ k} = 0,\nonumber 
    \\
    \kappa B+e\rho^{l} = 0,\label{NCGauss}
\end{gather}
where $B=\epsilon_{ij}F_{ij}$,  $E_{i}=F_{i0}$. Note, however,
that
$\rho^l$ and $\vec{j}\,{}^l$ denote here the {\it left density}
and {\it left current}, respectively,
\begin{gather*}
    \rho^{l}=\psi\star\bar\psi,
    \qquad
    \vec{j}\,{}^l=\frac{1}{2i}\left(\vec{D}\psi\star\bar\psi
    -\psi\star(\overline{\vec{D}\psi})\right).
\end{gather*}

These theories admit static, f\/inite-energy vortex solutions~\cite{NCFT}
which generalize those found before in ordinary CS theory~\cite{ZHK,JP}.

Are these theories Galilean invariant? At f\/irst sight, the
answer seems to be negative, and it has been indeed a widely
shared
view that Moyal f\/ield theory is inconsistent with Galilean
symmetry. The situation is more subtle, however. The conventional inf\/initesimal implementation of a
Galilean boost,
\begin{gather*}
\delta^0B=-t\vec{b}\cdot\vec{\nabla}B
\qquad\hbox{but}\qquad
\delta^0\rho^{l}=-\frac{\theta}{2}\vec{b}\times\vec{\nabla}\rho^l
-\vec{b}\cdot\vec{\nabla}\rho^l.
\end{gather*}
is indeed broken, as the Gauss constraint (\ref{NCGauss}) is
not preserved. Galilean symmetry can be restored taking into account the Moyal structure \cite{ExoFT}, namely
 considering  the {\it antifundamental representation}
\begin{gather*}
    \delta^{r}\psi=\psi\star(i\vb\cdot\vx)-t\vb\cdot\vnabla\psi
    =
    (i\vb\cdot\vx)\psi+\frac{\theta}{2}\vb\times\vnabla\psi
    -t\vb\cdot\vnabla\psi.
\end{gather*}
Observing that
\begin{gather*}
    \delta^r \psi=\delta^0 \psi+\frac{\theta}{2}\vb\times\vnabla\psi
\end{gather*}
we f\/ind that the $\theta$-terms cancel in $\delta^r\rho^l$,
leaving us with the homogeneous transformation law
\begin{gather*}
    \delta^{r}\rho^{l}=-t\vb\cdot\vnabla\rho^{l}.
\end{gather*}
Putting
$
\delta^{r}A_{\mu}=\delta^{0} A_{\mu},
$
so that $\delta^rB=\delta^0B$, the
Gauss constraint (\ref{NCGauss}) is right-invariant,
as are all the remaining equations.
The associated boost generator, calculated using the Noether
theorem, reads
\begin{gather}
    \vec{K}^r=t\vP-\int \vx \rho^r  d^2\vx,
    \label{rncboost}
\end{gather}
where
\begin{gather*}
    P_{i}=\int \frac{1}{2i}
    \big(\bar{\psi}\p_{i}\psi-(\overline{\p_{i}\psi})\psi\big) d^2\vx
    -\frac{\kappa}{2}\int \epsilon_{jk}A_{k}\p_{i}A_{j} d^2\vx
\end{gather*}
is the conserved momentum. The
conservation of (\ref{rncboost}) can also be checked directly, using the continuity
equation satisf\/ied by the right density, $\rho^r=\bar{\psi}\star\psi$.
At last, the boost components have the exotic commutation
relation
\begin{gather*}
    \big\{K_{i}, K_{j}\big\}=\epsilon_{ij}k,
    \qquad
    k\equiv
    -\theta \int \vert\psi\vert^2 d^2x.
\end{gather*}

Let us note, in conclusion, that Galilean symmetry
as established here makes it possible to produce
moving vortices by boosting the static solutions
constructed in~\cite{NCFT}, see \cite{moving}.

\section{Noncommutativity in 3 dimensions:\\ the semiclassical Bloch electron}\label{Bloch}

\subsection{The semiclassical model}

Around the same time and with no relation to the above
developments, a very similar theory has arisen in condensed matter
physics. For instance, applying a Berry-phase argument to a~Bloch
electron in a lattice,
 the  standard semiclassical equations~\cite{Mermin}
are modif\/ied by new terms~\cite{Niu},
generating purely quantum ef\/fects on  the mean values of the
electron's position and quasi-momentum ${\vr}$ and ${\vp}$,
respectively, which add to the force due to the momentum gradient
of the energy
band dispersion relation $\epsilon_n({\vp})$ and to the external (for instance, Lorentz) forces.
The  semiclassical approach allows several applications and generalizations,
 both from the physical \cite{Niu,AHE,SpinHall,Optical,RoVe}, and the mathematical \cite{Hreview,DHHMS,BlochHam,BeMo}
 side.

The clue is that the semiclassical model f\/its perfectly into Souriau's  general framework
 \cite{SSD} presented above.
  One starts with
a ``microscopic'' Hamiltonian operator $ {\hat H}\big[
{\hat\vr},{\hat\vp}, f( {\hat\vr}, t ) \big] $ for a~particle
(electron) for a periodic potential,
which is  adiabatically (in space-time) modif\/ied by a~perturbation
 $f$ (possibly an external f\/ield).
The  position/momentum operators  $\hat\vr$ and ${\hat
\vp}$ satisfy the Heisenberg algebra,
as usual. Moreover, for any constant~$f$, the Hamiltonian  ${\hat H}$ reduces to the usual one for  a
periodic crystal lattice.

The adiabatic features of the perturbation $f$ are expressed by
the inequalities $l_{\rm latt} \ll l_{w p} \ll l_{\rm mod}$, among  the
lattice constant length $l_{\rm latt}$, the wave-packet dispersion
length $l_{w p}$ and the modulation wave-length $l_{\rm mod}$.
Furthermore, the characteristic time scale $\hbar/\Delta E_{\rm gap}$
must be much smaller than the  typical time-scale of variations of~$f$.

The  f\/irst order truncation of the Hamiltonian
 around the instantaneous mean
position $\vr_c$,
\begin{gather*}
 {\hat H}\big[
{\hat\vr},{\hat\vp}, f({\hat\vr}, t ) \big] = {\hat H}_{(\vr_c,
t)} + {\hat W}_{(\vr_c, t)} ,\nonumber
\\
 {\hat W}_{(\vr_c, t)} = \frac 12 \big[ \p_f
{\hat H}\, \nabla_{\vr_c } f ( {\vr_c}, t ) \cdot ( {\hat
\vr} - \vr_c ) + {\rm h.c.} \big] , 
\end{gather*}
def\/ines a
quasi-static Hamiltonian ${\hat H}_{\lf \vr_c,t \rg}$, depending
on the ``slow'' parameters $c=\lf{\vr_c}, t\rg$.
${\hat H}_{\lf \vr_c,t \rg}$ is periodic
under~${\vec a}$~-- translations, and its eigenstates are Bloch the
waves. The latter are def\/ined,
for any f\/ixed
time $t$ and  $\vr_c$, by
\begin{alignat*}{3}
&  {\hat
H}_{\lf\vr_c,t\rg}|\psi_{\lf\vr_c,t\rg}^{n,\vq}\rangle  =
E_{\lf\vr_c,t\rg}^{n,\vq}|\psi_{\lf\vr_c,t\rg}^{n,\vq}\rangle,
\qquad && \langle \psi_{\lf\vr_c,t\rg}^{n ,\vq}
|\psi_{\lf\vr_c,t\rg}^{n',\vq \,'}\rangle  = \delta_{n,n'} \delta
\lf \vq-\vq\,'\rg, &
\\
& \langle \vr|\psi_{\lf\vr_c,t\rg}^{n ,\vq}\rangle  = e^{i\vq \cdot
\vr } u_{\lf\vr_c,t\rg}^{n ,\vq} \lf \vr \rg, \qquad &&
u_{\lf\vr_c,t\rg}^{n ,\vq} \lf \vr + {\vec a}\rg  = u_{\lf \vr_c,t
\rg}^{n ,\vq}\lf \vr \rg , & 
\end{alignat*}   where the energy eigenvalues
$E_{\lf\vr_c,t\rg}^{n ,\vq}$ are labeled by the band index, $n$,
and by the quasi-momentum $\vq$, restricted to the f\/irst Brillouin
zone (IBZ).
 We assume that the
time evolution of  ${\vr_c}$  closely follows the one obtained by
the exact integration of the Schr\"odinger equation and that the
eigenvalues $E_{\lf\vr_c,t\rg}^{n,\vq}$ form  well separated bands,
and that band jumping is forbidden. The  label~$n$ will be dropped in what follows.

A classical result by Karplus and Luttinger \cite{Lutti} says that
\begin{gather}
 \langle\psi_{\lf\vr_c,t
\rg}^{\vq}|\,{\hat\vr}\,|\psi_{\lf\vr_c,t\rg}^{\vq'} \rangle=\big[
i\nabla_{\vq}+\,\langle u_{\lf\vr_c,t\rg}^{\vq} \lf \vr
\rg|i\nabla_{\vq}\,u_{\lf\vr_c,t\rg}^{\vq \,} \lf \vr
\rg\rangle_{\rm cell} \big] \delta\lf\vq\,'-\vq\rg. \label{MatrixEl}
\end{gather}
That is,  the momentum  representation of ${\hat\vr}$ is
 \begin{gather*}
{\hat\vr}=i\nabla_{\vq}+\vec{\cA}\lf \vr_c,\vq ,t\rg
, \qquad \vec{\cA} = \langle u_{\lf
\vr_c,t\rg}^{\vq}|i\nabla_{\vq}\, u_{\lf\vr_c,t \rg}^{\vq}
\rangle_{\rm cell},
\end{gather*}
 where $\langle\cdot|\cdot\rangle_{\rm cell}$ is
the restriction of the scalar product to the unit cell with
periodic boundary conditions, and with  normalization factor $\lf
2 \pi \rg^3/V_{\rm cell}$. Then the quantity $\vec{\cA}\,{\lf \vr_c,
\vq ,t \rg}$ is interpreted as a $U(1)$ \emph{Berry connection}, whose
curvature appears in the commutation relations for  the
position operator components,
\begin{gather}
 \lq{\hat r}_j,{\hat r}_l \rq
= i\, \epsilon_{j l } \, \p_{q_j} {{\cA}}\,_l {\lf \vr_c, \vq ,t
\rg} = \Theta_{jl}\lf\vr_c,\vq,t\rg, \label{BerryTens}
\end{gather}
which converts the dynamics  of an ordinary particle in a periodic
background potential  into a~quantum mechanical system in a~\emph{non-commutative configuration space}~\cite{Szabo}. The antisymmetric tensor $\bTheta=(\Theta_{ij})$ generalizes in fact the scalar parameter $\theta$ of the
planar non-commutative theory.

 Its ef\/fects cannot be disregarded
 for the semiclassical motion of a wave-packet
\begin{gather*}
|\widetilde{\Psi}\lq \vr_c\lf t \rg, \vq_c\lf t \rg\rq \rangle =
\int_{\rm IBZ}  \Phi\lf \vq, t \rg | \psi_{\lf \vr_c,t \rg}^{\vq }
\rangle \; d  \vq,
\end{gather*}
 built by superimposing one-band Bloch waves with
a normalized amplitude $\Phi\lf \vq, t \rg $.
In fact, under the assumptions of small momentum dispersion,
$\Delta_q \ll  2\pi/ l_{\rm latt}$,  it can be proved that the
 mean packet-position is
  \begin{gather*}
  \vr_c \lf t
\rg =
   \langle \widetilde{\Psi} | \,
  {\hat {\vr}}\, |
  \widetilde{\Psi} \rangle \approx   -\nabla_{\vq_c}\,
  {\rm arg} \lq \Phi\lf \vq_c, t \rg \rq  +
  {\vec{\cA}}\,{\lf \vr_c, \vq_c ,t \rg },
 \end{gather*}
where the mean quasi-momentum,
\begin{gather}
 \vq_c \lf t \rg = \int_{_{IBZ}}
\vq \; | \Phi\lf \vq, t \rg |^2 \; d\, \vq,
\end{gather}
 has been
introduced. Then, the semiclassical description of the wave-packet
is reduced to   that of a  particle -- like  system in the $\lf
\vr_c ,   \vq_c \rg$ ``phase space'', the dynamics of which is
obtained by  minimizing the Schr\"odinger f\/ield  action
\begin{gather*}
 S =
\int_{t_1}^{t_2} \left\{ \frac{i}{2}
 \frac{\langle \Psi| \frac{d \Psi}{d t}\rangle -
  \langle \frac{d \Psi}{d t}| \Psi \rangle }
{\langle \Psi| \Psi \rangle } - \frac{\langle \Psi|{\hat H}| \Psi
\rangle }{\langle \Psi| \Psi \rangle}\right \} dt, 
\end{gather*}
 where  $\lf \vr_c \lf t \rg ,   \vq_c \lf t \rg\rg$
parametrize the wave-function~\cite{Kramer}.  This leads to an ``approximate  Lagrangian''
 for a point-like classical particle of the form~(\ref{15}),
 namely to
\begin{gather} L_{\rm app} =
  \dot{\vec r}{}_c  \cdot \lf \vq_{c} +
 \overrightarrow{{\cal R}}\lf \vr_c, \vq_c, t \rg\rg  +
  \dot{\vq}_{c}
 \cdot{\vec{\cA}}\,{\lf \vr_c, \vq_c, t \rg } +
 {\cal T}\lf \vr_c, \vq_c, t\rg  \nonumber\\
 \hphantom{L_{\rm app} =}{}  - {\cE}\lf \vr_c, \vq_c, t \rg  -
 \Delta{\cE}\lf \vr_c, \vq_c, t
 \rg, \label{Lagrapp}
 \end{gather}
where
\begin{gather*}
{\cal T}\lf\vr_c,\vq_c,t\rg = \langle u_{\lf \vr_c,t
\rg}^{\vq_c} | i \p_t \, u_{\lf\vr_c,t\rg}^{\vq_c} \rangle_{\rm cell},\nonumber
\\
\overrightarrow{{\cal R}}\lf\vr_c,\vq_c,t\rg  =  \langle
u_{\lf \vr_c,t \rg}^{\vq_c} |i  \nabla_{\vr_c} \,
u_{\lf\vr_c,t\rg}^{\vq_c} \rangle_{\rm cell},
\\
{\cE} =
 \langle \widetilde{\Psi} | {\hat H}_{\lf \vr_c,t \rg}
  |\widetilde{\Psi}\rangle,\qquad
\Delta{\cE} =
  \langle \widetilde{\Psi} |  {\hat W}_{\lf \vr_c,t \rg}  | \widetilde{\Psi}\rangle .\nonumber
\end{gather*}
Together with $\vec{\cA}$, the scalar ${\cal T}$ and the
vector f\/ield $\overrightarrow{{\cal R}}$ provide us with the
complete Berry connection on the entire  ``environmental parameter
space'' $\lf \vr_c,  \vq_c ,  t \rg$. The quantity ${\cE}$
expresses the potential energy felt by the wave packet in the
periodic environment and $\Delta{\cE}$  comes from the adiabatic
perturbations.

For slowly changing electromagnetic  potentials $\big(
\vA\lf\vr,t\rg, V_{el}\lf\vr,t\rg\big)$, the rather involved
expressions above take an elegant form~\cite{Niu},~-- but  the
Bloch eigenfunctions
 get a gauge-dependent phase modif\/ication $\approx e\vA(\vr_c, t)
\cdot\vr$.
In fact,  a change of phase has no
inf\/luence on the Berry connection because of  (\ref{MatrixEl}), so one can introduce the gauge
invariant kinetic momentum
\begin{gather*}
 \vk_c=\vq_c-e \vA(\vr_c,t)
\end{gather*}
and set again
\begin{gather*}
 {\vec{\cA}}\,{\lf \vr_c, \vq_c, t \rg }
={\vec{\cA}}\,(\vk_c), \\
\overrightarrow{{\cal R}} \simeq -e\,\nabla_{\vr_c}\big(
{\vA}\,(\vr_c,t)\cdot\vr\big)\Big|_{\vr=\vr_c},\qquad {\cal T}
\simeq-e\,\p_t {{\vA}}\,{(\vr_c,t)\cdot\vr_c },
\\
 {\cE}={\cE}_0(\vk_c) +e\,V_{el}(\vr_c, t),
\qquad  \Delta{\cE}= - {\vec M}( \vk_c, t)\cdot\vB (\vr_c, t),
\end{gather*}
 where
 \[ {\vec M}(\vk_c, t)=-\frac{e}{2 m}
\langle\widetilde{\Psi}|{\hat{\vec L}}|\widetilde{\Psi}\rangle
\] is the mean magnetic moment
of the wave-packet. $\vB (\vr_c,t)$ and ${\vec E} (\vr_c,t)$ are
def\/ined as usual from the  mean values of the potentials. Dropping the label $c$, and putting
\begin{gather*}
\Theta_i=\half \epsilon_{i j k}
\Theta_{jk},
\end{gather*}
the generalized semiclassical equations of
motion  are
\begin{gather}
 \dot{\vr}   =   \nabla_{\vk}\big[\cE_0(\vk)-{\vec
M}(\vk,t) \cdot {\vec B}(\vr , t )\big] - {\dot{\vk}}\times
{\vTheta}(\vk), \nonumber\\
\dot{\vk}   =   - e \big( {\dot{\vr} }\times {\vB}(\vr, t ) +
{\vec E}(\vr, t)\big) + \nabla_{{\vr}  }\big( {\vec M}( \vk, t)
\cdot {\vB}(\vr,t) \big), \label{EqMot}
\end{gather}
 further
conf\/irming the idea of the non-commutativity parameter,
 now a function of momentum-space variables, is in fact
  a Berry phase ef\/fect.

Notice, here that the semiclassical procedure has consistently
``averaged'' on the gauge degrees of freedom at local scales of
order $ \sim l_{wp}$, but  the f\/inal model still possesses the
same gauge invariant character  as a point-like particle
interacting with an external e.m.~f\/ield.

Then, for the electronic  wave-packet semiclassically described by
(\ref{EqMot}), one can adapt the symplectic techniques described
in the previous sections and it can be used for a
 Hamiltonian formulation.

\subsection{Hamiltonian structure}

Comparing the system (\ref{EqMot}) with the previous ones in
(\ref{DHeqmot}) or (\ref{815}), one recognizes a general common
structure.
 The nice
group-geometrical symmetry properties of the 2D Galilei group,
which
partially motivated the present research, are broken in general.
However, the unifying framework for such dif\/ferential systems is
provided by the same ideology adopted in Sections~\ref{Sec2} and~\ref{Sec6}, i.e.\
writing them as the kernel of a postulated  anti-symmetric,
closed, constant-rank Lagrange--Souriau
 2-form $\sigma$  of the form
\begin{gather}
 \sigma = \lq \lf 1 - Q_{i}\rg dq_i - e  E_i  dt \rq \wedge
\lf dr_i - g_i  dt \rg
\nonumber\\
\phantom{\sigma =}{}  + \frac 12 e   \epsilon_{ijk} B_k  dr_i \wedge dr_j
 +\frac 12 \epsilon_{i j k} \Theta_k   dq_i \wedge dq_j +  Q_0
\epsilon_{i j}  dr_i \wedge dq_j , \label{Lag2D}
 \end{gather}
where the  Souriau's prescription   to explicitly include the
electromagnetic contributions has been used. The vector
f\/ields
 ${\vec g}$, ${\vTheta}$, ${\vec Q}$ and the scalar functions $ Q_0$
may depend on all independent variables $\lf  \vr, \vq, t \rg $. Notice that (\ref{Lag2D}) only contains  ``forces'', i.e.\
 gauge invariant quantities. Moreover, the so called ``Maxwell  principle'' \cite{SSD}, i.e.\ the closure relation
$d \sigma = 0$, implies a set of integrability conditions for
functions involved, which reduce to the usual  Maxwell equations
for  $\big( {\vec E} ,   {\vec B} \big)$, when the new extra f\/ields
are set to constants. Even in this case, and in 2 space
dimensions, the resulting equations are non-trivial, coinciding
for instance with~(\ref{DHeqmot}) after the identif\/ications $r_i
\rightarrow x_i$, $g_i \rightarrow q_i/m $, $\theta_3 \rightarrow
\theta$ and $Q_i \equiv 0$.

We note (like in Section~\ref{Sec2}) that a model def\/ined by the 2-form
(\ref{Lag2D}) may not possess a globally def\/ined conf\/iguration
space Lagrangian.  This makes the value of the
 semiclassical Lagrangian~(\ref{Lagrapp}) questionable.  If it is assumed valid at least
locally, the physical meaning of the coef\/f\/icients appearing in
(\ref{Lag2D}) can be deduced, via exterior derivative, from  the
Cartan 1-form
\begin{gather*}
 \lambda = \big( \vq +
 \overrightarrow{{\cal R}} \big) \cdot d \vr + {\vec{\cA}}  \cdot d \vq + \lf {\cal T}  - {\cE}  - \Delta{\cE} \rg  dt.
\end{gather*}

Thus, the most general equations of  motion deriving from
(\ref{Lagrapp}) (or equivalently from (\ref{Lag2D})) are
 \begin{gather}
  \lf 1 + \bXi \rg  \dot{\vr}  +
\Theta   {\dot{\vq  }}    =
 \nabla_{\vq  } \lq \cE
+ \Delta{\cE} - \cT \rq + \p_t {\vec \cA}, \nonumber\\
  X   {\dot{\vr}
} + \lf 1 + \bXi  \rg \dot{\vq}     =   -\nabla_{\vr  } \lq \cE +
\Delta{\cE} - \cT \rq  -\p_t {\vec \cR},
\label{eqmotgen}
\end{gather}
 where the antisymmetric
matrices $\bXi=(\Xi_{ij})$ and $\bX=(X_{ij})$ have  elements
\begin{gather}
 \bXi_{i j} = \p_{r_i}
\cA_j - \p_{ q_j} \cR_i ,  \qquad  X_{ij} =
 \p_{r_i}\cR_j - \p_{r_j}\cR_i .
 \label{altriTens}
 \end{gather}
The dynamical system (\ref{eqmotgen}) is def\/ined on the tangent
manifold of the conf\/iguration space, endowed with generalized
coordinates $\vec{\xi} = \lf \vr  , \vq \rg$. But, when
$\p_t \vec{\cA} = \p_t \vec{\cR} \equiv 0 $, the  rearrangement $
\sigma=\omega-dH\wedge dt $ of the terms in (\ref{Lag2D}) is
possible, introducing the symplectic 2-form {\samepage
\begin{gather*}
 \omega = \lf
\delta_{i,j}+ \Xi_{ij} \rg d r_{i} \wedge d q_{j} +
\frac{1}{2} \lq X_{i j}  d q_{i} \wedge d q_{j} - \Theta_{i
j}  d r_{i} \wedge d r_{j} \rq 
\end{gather*}
 and the Hamiltonian function $ {\cal H} = \cE + \Delta{\cE} - \cT $.}

 Actually, the closure of $\sigma$ implies that, $d\omega=0$, for
$\omega$.
  Equivalently, the  set of dif\/ferential constraints
\begin{alignat}{3}
&  \varepsilon_{i j k}   \p_{q_i} \Theta_{j k} = 0, \qquad &&
 \varepsilon_{i j k}   \p_{r_i} X_{j k} = 0 , &\nonumber\\
&  \p_{q_j}   \Xi_{i j} = -   \p_{r_j}   \Theta_{i j} ,
 \qquad &&  \p_{r_j}   \Xi_{i j} =    \p_{q_j}   X_{i j}, &
 \label{Closure}
 \\
 & \lf 1-\delta_{h  k}\rg \varepsilon_{k  i  j}  \p_{q_k}
 \Xi_{i j} =  \varepsilon_{h i j}  \p_{r_h}
 \Theta_{i j},\qquad && \lf 1-\delta_{h k}\rg \varepsilon_{k i j}  \p_{r_k}
 \Xi_{i j} = - \varepsilon_{h i j}  \p_{q_h}
 X_{i j},\nonumber&
 \end{alignat}
which, however, are automatically satisf\/ied, because of the antisymmetry and the di\-f\/fe\-ren\-tia\-bi\-li\-ty properties
of the  tensors $\bTheta$, $\bXi$ and $\bX$  def\/ined in
(\ref{BerryTens}) and (\ref{altriTens}).
Thus, for  non degenerate $\omega = \omega_{\alpha \beta}   d
\xi_{\alpha} \wedge d \xi_{\beta}$, Poisson brackets,
\begin{gather*}
\lgr f, g
\rgr = \omega^{\alpha\beta} \p_{\alpha} f \p_{\beta} g
\end{gather*}
 can be
def\/ined for any pair of functions $f( \vec{\xi} )$ and $g( \vec{\xi} )$,  where $\omega^{\alpha\gamma}\omega_{\gamma
\beta} = \delta^{\alpha}_{\beta}$ is the inverse of the symplectic
matrix \cite{SSD, Marmo}. Thus, the equations (\ref{eqmotgen}) take the usual Hamiltonian form
$\dot{\xi}_\alpha=\{\xi_\alpha,{\cal H}\}.$ In the present case
$\lf \omega_{\alpha\beta} \rg$ is a real symplectic $6\times 6$
matrix, which  is non degenerate when
\begin{gather*}
\sqrt{\det\lf
\omega_{\alpha\beta}\rg} = 1-\frac{1}{2}{\rm Tr}
 \lf \bXi^2 + \bX \lf {\bf 1} +
2   \bXi \rg \bTheta \rg \neq 0.
\end{gather*}
 Such a factor
generalizes the denominators present in the Poisson brackets
(\ref{exocommrel}), (\ref{PB4}) or (\ref{PB8}). Moreover, it
crucially appears in the expression of the invariant phase-space
volume, ensuring the validity of the Liouville theorem \cite{Xiao,BlochHam}.

As special example, we deal with  only momentum (gauge invariant)
dependent Berry curvature ${\vTheta}\lf \vq \rg$
 which is to be divergence-free
according to the f\/irst equation in  (\ref{Closure}). That
condition can be satisf\/ied, except in one point, e.g., by
 a {\it monopole in ${\vq}$-space},
\begin{gather}
{\vTheta} = g \frac{\vq} {q^3}  ,
\label{pmonop}
\end{gather}
which is indeed the only possibility consistent with the spherical
symmetry and the canonical relations $\{x_i,q_j\}=\delta_{ij}$~\cite{BeMo}. The expression~(\ref{pmonop}) appears to be
consistent, at least qualitatively, with the data reported in~\cite{AHE} and in Spin  Hall Ef\/fects \cite{SpinHall}.

In  absence of a magnetic f\/ield and taking, for  simplicity, the
energy band $\epsilon_n({\vq})$ to be parabolic, the equations
(\ref{EqMot}) for  become
\begin{gather*}
\dot{\vr}={\vq} +\frac{e   g}{q^3}{\vec E}\times{\vq}, \qquad
\dot{\vq} = - e  {\vec E}. 
\end{gather*}
The anomalous term shifts the velocity and deviates, hence, the
particle's trajectory perpendi\-cu\-larly to the electric f\/ield, just
like in the anomalous Hall ef\/fect, see~\cite{AHE}.

A similar pattern arises in optics \cite{Optical,OptiHall,SpinOptics,OpticReview}:
to f\/irst order in the gradient of the refractive index $n$, spinning
light is approximately described by the equations
\begin{gather*}
\dot{\vr}\approx{\vp}-\frac{s}{\omega}\,{\rm grad
}\left(\frac{1}{n}\right)\times{\vp}, \qquad \dot{\vp}\approx
-n^3\omega^2{\rm grad }\left(\frac{1}{n}\right), 
\end{gather*}
where $s$ denotes the photon's spin. In the f\/irst relation we
recognize, once again, an anomalous velocity relation of the type~(\ref{EqMot}). The new term makes the light's trajectory deviate
from that predicted in ordinary geometrical optics, giving rise to
the ``optical Magnus ef\/fect''~\cite{Optical}. A~manifestation of
this is the displacement of the light ray perpendicularly to the
plane of incidence at the interface of two media with dif\/ferent
refraction index: this is the ``Optical Hall Ef\/fect''
\cite{OptiHall,SpinOptics,OpticReview}.

Another nice illustration is provided by the non-commutative Kepler problem \cite{RoVe}. Choosing
the non-commutative vector $\vTheta$ in the vertical direction,
\begin{gather*}
\Theta_i=\theta\delta_{iz}
\end{gather*}
the 3D problem reduces to the
``exotic'' model presented in Section~\ref{Sec2}.
Then the authors of~\cite{RoVe} show that, for the Kepler potential $V\propto r^{-1}$ the perturbation due to non-commutativity induces the precession of the perihelion point of planetary orbit.

As yet another example, we would like to mention the recent work \cite{ZHN}, in which it is shown that a  particle with ``monopole-type''
noncommutativity (\ref{pmonop}) admits  a conserved Runge--Lenz vector,
namely
\begin{gather*}
\vec{K}=\vr\times\vec{J}-\alpha\frac{\vq}{q}
 ,
\end{gather*}
provided the Hamiltonian
is
\begin{gather*}
H=\frac{\vr\,{}^2}{2}+\frac{g^2}{2q^2}+
\frac{\alpha}{q}  .
\end{gather*}
Note that $\vq$ here is the momentum: the ``monopole'' is in ``dual space''.
Let us observe that this expression is reminiscent of the Chern--Simons mechanics'' \cite{DJT} in that it has no mass term. The associated the equations of motion read,
\begin{gather*}
\dot{\vr}=-\left(\frac{g^2}{q^4}+\frac{\alpha}{q^3}\right)\vq
+g \frac{\vq\times{\vr}}{q^3},
\qquad
\dot{\vq}=-{\vr} .
\end{gather*}

The Kepler-type dynamical symmetry
then allows one to show that the classical motions follow (arcs of)
oblique ellipses~\cite{ZHN}.

\goodbreak
\subsection*{Acknowledgments}
 This review also includes results obtained jointly  with
C.~Duval, Z.~Horv\'ath, J.~Lukierski, M.~Plyushchay
and W.J.~Zakrzewski,
 to whom we express our indebtedness.
L.M.\ thanks the INFN - Sezione of Lecce for partial f\/inancial support under the project LE41.

\pdfbookmark[1]{References}{ref}
 \LastPageEnding

\end{document}